\documentclass[
reprint,
superscriptaddress,
super,
]{revtex4-2}

\usepackage{graphicx}
\usepackage{dcolumn}
\usepackage{bm}

\usepackage[utf8]{inputenc}
\usepackage[T1]{fontenc}
\usepackage{titlesec}
\titleformat{\section}{\normalfont\large\bfseries}{}{0em}{}
\titlespacing{\section}{0pt}{10pt}{0pt}

\usepackage{siunitx, xcolor}
\usepackage[version=4]{mhchem}

\usepackage{amsmath}
\usepackage{eucal}
\DeclareMathAlphabet{\mathcal}{OMS}{cmsy}{m}{n}
\DeclareSymbolFont{largesymbols}{OMX}{cmex}{m}{n}
\usepackage{mathrsfs}
\usepackage{epstopdf}
\usepackage[colorlinks=true, letterpaper=ture, pdfstartview=FitV, linkcolor=blue, citecolor=blue, urlcolor=blue]{hyperref}
\usepackage{dcolumn}

\makeatletter
\renewcommand{\cite}[1]{\textsuperscript{\citenum{#1}}}
\makeatother

\begin{document} 
	
	\title{All-electric control of skyrmion-bimeron transition in van der Waals heterostructures}%
	
	\author{Lan Bo}%
	\thanks{These authors contributed equally to this work}
	\affiliation{School of Science and Engineering, The Chinese University of Hong Kong, Shenzhen, Guangdong 518172, China}
	\affiliation{Department of Applied Physics, Waseda University, Okubo, Shinjuku-ku, Tokyo 169-8555, Japan}
	\author{Songli Dai}%
	\thanks{These authors contributed equally to this work}
	\affiliation{Institute of Advanced Optoelectronic Materials and Technology, College of Big Data and Information Engineering, Guizhou University, Guiyang 550025, China}
	\affiliation{School of Science and Engineering, The Chinese University of Hong Kong, Shenzhen, Guangdong 518172, China}
	\author{Xichao Zhang}%
	\affiliation{Department of Applied Physics, Waseda University, Okubo, Shinjuku-ku, Tokyo 169-8555, Japan}
	\author{Masahito Mochizuki}%
	\affiliation{Department of Applied Physics, Waseda University, Okubo, Shinjuku-ku, Tokyo 169-8555, Japan}
	\author{Xiaohong Xu}
	\affiliation{School of Chemistry and Materials Science of Shanxi Normal University \& Key Laboratory of Magnetic Molecules and Magnetic Information Materials of Ministry of Education, Taiyuan 030031, China } 
	\author{Zean Tian}%
	\email[Corresponding E-mail: ]{tianzean@hnu.edu.cn}
	\affiliation{College of Computer Science and Electronic Engineering, Hunan University, Changsha 410082, China}
	\affiliation{Institute of Advanced Optoelectronic Materials and Technology, College of Big Data and Information Engineering, Guizhou University, Guiyang 550025, China}
	\author{Yan Zhou}%
	\email[Corresponding E-mail: ]{zhouyan@cuhk.edu.cn}
	\affiliation{School of Science and Engineering, The Chinese University of Hong Kong, Shenzhen, Guangdong 518172, China}


	\begin{abstract}
		Two-dimensional van der Waals materials offer a versatile platform for manipulating atomic-scale topological spin textures. In this study, using first-principles and micromagnetic calculations, we demonstrate a reversible transition between magnetic skyrmions and bimerons in a MoTeI/In$_2$Se$_3$ multiferroic heterostructure. The physical origin lies in the reorientation of the easy axis of magnetic anisotropy, triggered by the reversal of ferroelectric polarization. We show that the transition operates effectively under both static and dynamic conditions, exhibiting remarkable stability and flexibility. Notably, this transition can be achieved entirely through electric control, without requiring any external magnetic field. Furthermore, we propose a binary encoding scheme based on the skyrmion-bimeron transition, presenting a promising path toward energy-efficient spintronic applications.
		
	\end{abstract}
	
	\maketitle

\begin{figure*}[t]
		\includegraphics[width=1\linewidth]{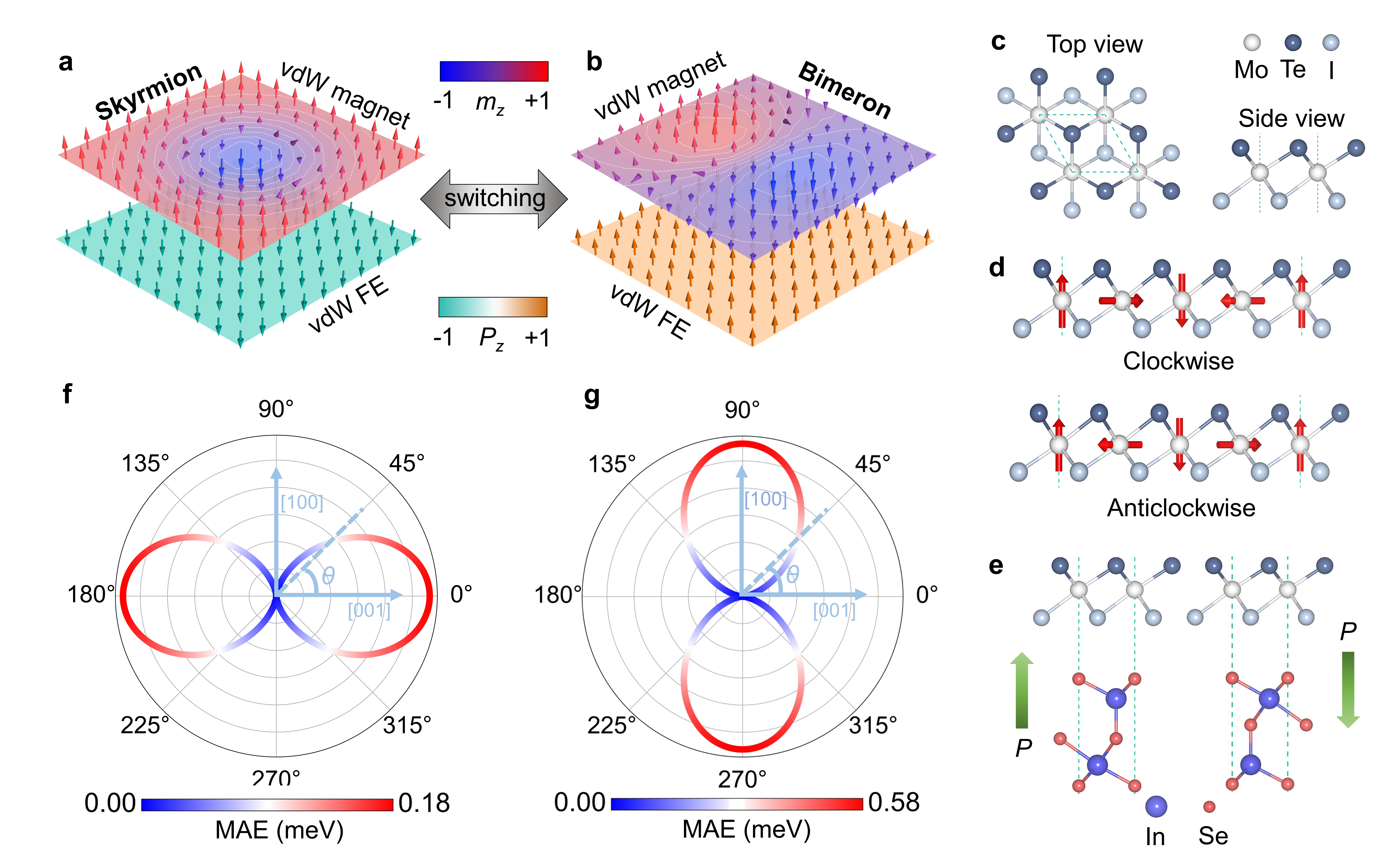}
		\caption{\textbf{Structural and magnetic properties of MoTeI monolayer and MoTeI/In$_2$Se$_3$ heterostructure.} Schematic diagram of \textbf{a} skyrmion and \textbf{b} bimeron switching in a 2D vdW magnetic layer (top) controlled by ferroelectric polarization reversal (bottom). \textbf{c} Top and side views of the atomic structure of the MoTeI monolayer. \textbf{d} Two spin configurations with opposite chirality, used to determine the in-plane DMI parameters, with red arrows indicating the spin orientations. \textbf{e} Side view of the MoTeI/In$_2$Se$_3$ heterostructure, depicting the two distinct polarization states, labeled P$\uparrow$ (left panel) and P$\downarrow$ (right panel). Dependence of magnetic anisotropy energy (MAE) on the polar angle $\theta$ in the \textbf{f} P$\uparrow$ and \textbf{g} P$\downarrow$ configurations for the MoTeI/In$_2$Se$_3$ heterostructure, respectively.
		}\label{1} 
	\end{figure*}

	Topological spin textures, serving as stable information carriers, have gained significant attention due to their potential in spintronic applications\cite{finocchio2016magnetic,kang2016skyrmion,fert2017magnetic,everschor2018perspective,tokura2020magnetic,back20202020,bogdanov2020physical,dieny2020opportunities,zhang2020skyrmion,bo2022micromagnetic}. Among these, magnetic skyrmions, as shown in Fig.~\ref{1}a, stand out due to their topological protection and energy-efficient manipulation via electric currents\cite{roessler2006spontaneous,nagaosa2013topological,mochizuki2015dynamical,jiang2017skyrmions,kanazawa2017noncentrosymmetric}. This renders them highly attractive for next-generation classical and quantum spintronic devices\cite{psaroudaki2021skyrmion,psaroudaki2023skyrmion,xia2023universal}. Similarly, bimerons, consisting of meron-antimeron pairs with opposite spin polarities (Fig.~\ref{1}b), can be viewed as the in-plane counterpart of skyrmions\cite{gobel2019magnetic,kim2019dynamics,li2020bimeron,sun2020controlling,shen2020dynamics,zhang2020static,liang2023bidirectional}. 
	
	The synthetic versatility and structural robustness of two-dimensional (2D) van der Waals (vdW) materials create a unique platform for exploring atomic-level spin textures, driven by the demand for miniaturization in modern electronics. Since 2019, magnetic skyrmions have been observed in 2D vdW magnets represented by Cr$_{2}$Ge$_{2}$Te$_{6}$\cite{han2019topological} and Fe$_{3}$GeTe$_{2}$\cite{ding2019observation}, as well as in their heterostructures\cite{park2021neel,wu2022van}. In addition, extensive theoretical and computational investigations have been conducted to predict further 2D vdW materials that can host skyrmions\cite{powalla2023seeding}. Specifically, 2D Janus monolayers, characterized by their intrinsic broken inversion symmetry, exhibit a pronounced Dzyaloshinskii–Moriya interaction (DMI) that stabilizes nonlinear spin textures\cite{liang2020very}. This potential becomes particularly promising when combined with 2D ferroelectric (FE) materials\cite{osada2019rise,guan2020recent,wu2021two} due to the emergence of spin-orbit coupling (SOC). Additionally, the electron reconstruction at the heterointerface provides a nonvolatile electric field, which further mediates magnetic interactions at the atomic scale\cite{wang2018ferroelectrically,wang2020electric,wang2020ferroelectric,ba2021electric}. For example, the magnetic anisotropy interaction can be significantly affected by reversing the FE polarization due to the proximity effect\cite{wang2024switching}. This enables control of topological spin textures using an electric field, offering the advantage of low energy dissipation\cite{srivastava2018large,ma2018electric}.
	
	Traditionally, spintronic devices have emphasized a single type of topological textures\cite{finocchio2016magnetic,kang2016skyrmion,fert2017magnetic}—either skyrmions or bimerons—determined by material properties. However, there has been a growing focus on their physical correlations\cite{araujo2020typical,silva2022skyrmion,yu2024spontaneous,li2024stability,zhang2021dynamic,sun2021manipulation,ohara2022reversible,castro2023skyrmion,yang2024strain,lei2025ferroelectric,zhang2024bimerons}, suggesting that integrating different topological spin textures could enhance storage and computing architectures, allowing multiple spintronic components to collaboratively execute complex tasks\cite{gobel2021beyond}. Topological spin textures can be described using homotopy theory\cite{mermin1979topological} and characterized by the topological charge\cite{nagaosa2013topological} $Q=(4\pi)^{-1}\iint{\bf m}\cdot \left({\partial_x {\bf m}} \times {\partial_y {\bf m}}\right) \, d^2r$, with ${\bf m}$ being the normalized magnetization vector. For a skyrmion, the boundary conditions are defined as ${\bf m}(r \!\rightarrow\!\pm\infty)\!\rightarrow\!(0,0,\pm1) $, and for a bimeron they are ${\bf m}(r \!\rightarrow\!\pm\infty)\!\rightarrow\!(0,\pm1,0) $\cite{araujo2020typical}. Thus, in an isotropic theoretical model, both skyrmions and bimerons exhibit identical nontrivial topological properties with $Q=\pm1$ and cannot be deformed continuously into a ground ferromagnetic state\cite{araujo2020typical}. However, the excitation of specific spin textures can be influenced by factors such as anisotropies or external fields that favor either out-of-plane or in-plane spin orientations\cite{kharkov2017bound}. Although recent studies have explored skyrmion-bimeron conversions\cite{zhang2021dynamic,sun2021manipulation,ohara2022reversible,castro2023skyrmion,yang2024strain,lei2025ferroelectric}, most efforts have focused on metastable scenarios, or relied on pre-patterned anisotropy profiles. The dynamic transitions within transport processes under realistic conditions remains largely unexplored. 
	
	In this work, we address this gap by conducting systematic first-principles calculations and micromagnetic simulations to investigate the skyrmion-bimeron transition. The transition is realized through all-electric control and validated within the dynamic motion process, which links the transformation with the widely studied skyrmion-based racetrack memory. The system are 2D vdW heterostructures, composed of Janus monolayer MoTeI and FE monolayer In$_2$Se$_3$, as shown in Fig.~\ref{1}a, b. Through modulation of the ferroelectric polarization in In$_2$Se$_3$, we achieve a controllable reorientation of the magnetic anisotropy in MoTeI, switching between in-plane and out-of-plane directions, consequently inducing changes in topological spin textures. We also demonstrate the stability and flexibility of this dynamic transformation and, based on this mechanism, design a racetrack memory-like binary encoder.
	
	\begin{figure*}[t]
		\includegraphics[width=1\linewidth]{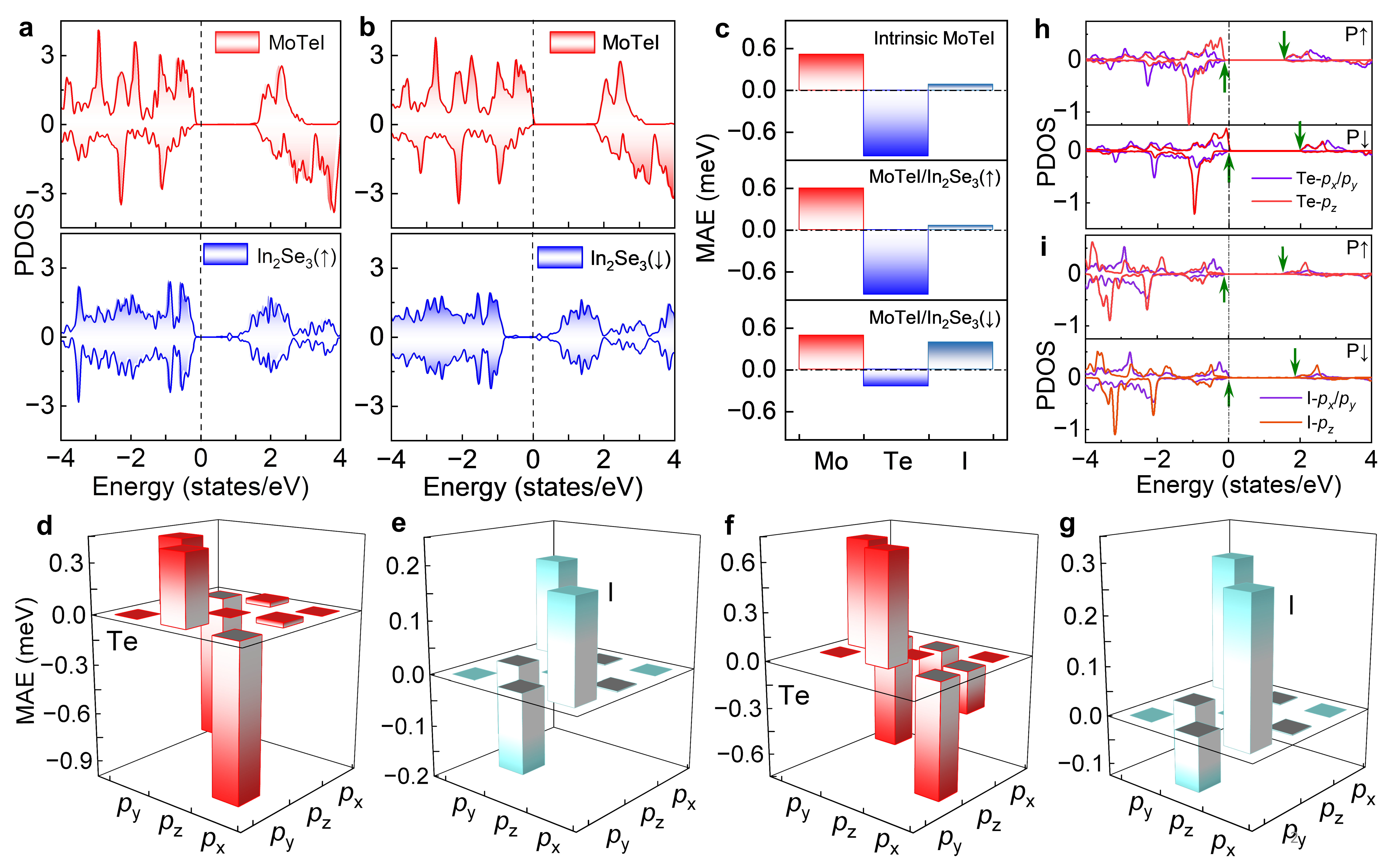}
		\caption{\textbf{Ferroelectric modulation of MAE in MoTeI/In$_2$Se$_3$ heterostructure.} Layer-resolved partial density of states (PDOS) of the MoTeI/In$_2$Se$_3$ heterostructure in the \textbf{a} P$\uparrow$ and \textbf{b} P$\downarrow$ configurations. \textbf{c} Atomic-resolved MAE comparison between the intrinsic MoTeI monolayer and the polarized MoTeI/In$_2$Se$_3$ heterostructure. Orbital-resolved MAE contributions from Te and I atoms in the MoTeI/In$_2$Se$_3$ heterostructure for both \textbf{d}, \textbf{e} P$\uparrow$ and \textbf{f}, \textbf{g} P$\downarrow$ configurations. PDOS of \textbf{h} Te and \textbf{i} I atoms within the MoTeI/In$_2$Se$_3$ heterostructure. The green arrows indicate the spin-down unoccupied states of the $p_x / p_y$ orbitals and the spin-up occupied states of the $p_y / p_x$ orbitals.
		}\label{2} 
	\end{figure*}
	
	\section*{Results}
	{\textbf{Electrical control of magnetic anisotropy in multiferroic MoTeI/In$_2$Se$_3$.}}
	We started by considering a MoTeI monolayer, whose crystal structure is displayed in Figs.~\ref{1}c, d with both top and side views, featuring a hexagonal lattice in $p3m1$ space group. This Janus structure is characterized by a stacking arrangement that arises from the misalignment of the Te, Mo and I layers, with an optimized lattice constant being 4.103 Å. The phonon spectrum (see Supplementary Fig. 1a) exhibits no imaginary frequencies, indicating the dynamic stability of the MoTeI monolayer. The total energy shows negligible fluctuations during the \textit{ab intio} molecular dynamics (AIMD) simulations at 300 K over 3 ps, with the MoTeI monolayer maintaining structural integrity (see Supplementary Fig. 1b), confirming its excellent thermal stability. The spin Hamiltonian for the MoTeI monolayer is expressed as
	\begin{equation}
		\mathcal{H} = -J \sum_{i, j} {\bf s}_i\cdot{\bf s}_j - {\bf D}_{i j} \sum_{i, j} {\bf s}_i \times {\bf s}_j - K \sum_i ({\bf s}_i\cdot\hat{\textbf{z}})^2
		,
	\end{equation}
	where ${\bf s}_i$ (${\bf s}_j$) is the spin of the $i$th ($j$th) Mo atom, $J$ is the exchange coupling constant, ${\bf D}_{ij}=(D_x,D_y,D_z)$ denotes the DMI vector, and $K$ is the single ion anisotropy constant. The SOC is included in the calculations of $K$ and ${\bf D}_{ij}$. According to Moriya's symmetry rules, the reflection planes intersect at the midpoint of bonds between adjacent Mo atoms, orienting each DMI vector perpendicular to the bonding axis. ${\bf D}_{ij}$ is defined as ${\bf D}_{ij}=d_\parallel (\hat{\textbf{z}}\times\hat{\textbf{u}}_{ij}) + d_\perp\hat{\textbf{z}}$, where $\hat{\textbf{u}}_{ij}$ is the vector connecting ${\bf s}_i$ and ${\bf s}_j$, and $d_\parallel$ and $d_\perp$ are the in-plane and out-of-plane DMI components, respectively. Since the magnitude of $d_\perp$ is small and its influence on spin textures is negligible in 2D systems, it is omitted in the subsequent calculations. The in-plane component $d_\parallel$ is determined by the chirality-dependent total energy difference approach, based on two spin configurations illustrated in Fig.~\ref{1}d. The exchange coupling constant $J$ was accurately calculated using a $4\times4\times1$ supercell through the four-state method\cite{xiang2011predicting}. Calculations reveal that the MoTeI monolayer serves as a 2D ferromagnet with an interlayer ferromagnetic coupling of 18.95 meV, an in-plane magnetic anisotropy energy (MAE) of -0.21 meV per unit cell, and an in-plane DMI component of 1.24 meV. In addition, MoTeI is identified as an indirect semiconductor with a band gap of 1.72 meV, as illustrated in Supplementary Fig. 2.
	
	To enable non-volatile modulation of the spin Hamiltonian in MoTeI, a multiferroic heterostructure was constructed by vertically stacking the MoTeI monolayer with 2D FE material In$_2$Se$_3$, as illustrated in Fig.~\ref{1}e. The optimized lattice constant of In$_2$Se$_3$ is 4.102 Å, consistent with previous experimental and theoretical studies\cite{ding2017prediction,cui2018intercorrelated}, and the lattice mismatch of approximately 0.048\% results in negligible strain effects in the heterostructure. Given the two stable polarization states (P$\uparrow$ and P$\downarrow$) of In$_2$Se$_3$, we explored six representative high-symmetry geometric stacking configurations, labeled type-I to type-VI (see Supplementary Fig. 3). To determine the most stable stacking mode among the configurations, we calculated the corresponding binding energy $E_{\rm b} = E_{\rm {MoTeI/In_2Se_3}}-E_{\rm {MoTeI}}-E_{\rm {In_2Se_3}}$. As shown in Supplementary Table I, type-I exhibits the lowest binding energy across both polarization states, making it the most stable stacking configuration. Thus, type-I stacking configuration is taken as our model system. We also evaluated the potential for realizing ferroelectricity in MoTeI/In$_2$Se$_3$ heterostructures. As shown in Supplementary Fig. 4, the transition pathway between P$\uparrow$ and P$\downarrow$ states exhibits an energy barrier of 0.71 eV per unit cell, similar to that of conventional ferroelectrics, indicating typical ferroelectricity in this system.
	
	We next examined the magnetic properties of the MoTeI/In$_2$Se$_3$ heterostructure. The magnetic parameters  for both P$\uparrow$ and P$\downarrow$ polarization states, calculated using density functional theory (DFT), are listed in Supplementary Table II, which show no significant variation in $J$ and $d_\parallel$ between these states. Figures~\ref{1}f, g show the dependence of MAE on the polar angle for both P$\uparrow$ and P$\downarrow$ polarization states. As the polarization switches from P$\uparrow$ to P$\downarrow$, the easy magnetization axis of the MoTeI monolayer shifts from in-plane (MAE of $-0.18$ meV per unit cell) to out-of-plane (MAE of 0.58 meV per unit cell). We also examined the influence of the ferroelectric layer on the electronic structure of MoTeI. Figures 2a,b present the layer-resolved partial density of states (PDOS) of the MoTeI/In$_2$Se$_3$ heterostructure under the P$\uparrow$ and P$\downarrow$ configurations. For the P$\uparrow$ configuration, MoTeI remains semiconducting, while for P$\downarrow$, the valence band maximum in the spin-up channel shifts upward, giving rise to a half-metallic state in MoTeI. This transition is electrostatically induced by switching the polarization of the underlying In$_2$Se$_3$ layer, and does not involve charge injection or transport through MoTeI. The resulting modulation of spin-resolved band structure provides a non-volatile means of controlling local magnetic configurations.
	
	To elucidate the origins of MAE in the MoTeI monolayer and MoTeI\textbf{/}In$_2$Se$_3$ heterostructures, we calculated the atom- and orbital-resolved MAEs for different configurations. As shown in Fig.~\ref{2}c, the results reveal that in both polarization states, Mo, Te, and I atoms are the primary contributors to MAE, while contributions from In and Se atoms are negligible. For both the intrinsic MoTeI monolayer and MoTeI\textbf{/}In$_2$Se$_3$ under P$\uparrow$ polarization, the overall in-plane MAE is primarily determined by substantial in-plane contribution from Te atoms, along with smaller out-of-plane contributions from Mo and I atoms. Under P$\downarrow$ polarization, the Te atoms’ in-plane contribution decreases considerably, whereas the I atoms’ out-of-plane contribution increases markedly. Consequently, the reorientation of easy magnetization axis from in-plane to out-of-plane is primarily attributed to the MAE contributions of Te and I atoms. Figures~\ref{2}d-g illustrate the contributions of the coupling between the $p$-orbital components $p_x$, $p_y$, and $p_z$ of Te and I atoms to the MAE in both P$\uparrow$ and P$\downarrow$ states. The atomic-resolved MAE of Te and I atoms are mainly driven by the coupling between the $p_x / p_y$ and $p_y / p_z$ orbitals. The observed reduction in in-plane MAE for Te atoms and out-of-plane MAE for I atoms can be primarily attributed to the coupling of $p_x / p_y$ orbitals. According to second-order perturbation theory \cite{wang1993first,yang2017strain}, the MAE can be expressed as 
	\begin{align}
		\text{MAE} &= \sum_{\sigma,\sigma^{\prime}} \big( 2 \delta_{\sigma,\sigma^{\prime}} - 1 \big) \xi^2  \\
		&\times \sum_{o^\sigma,u^{\sigma^{\prime}}} 
		\frac{\big| \langle o^\sigma | L_z | u^{\sigma^{\prime}} \rangle \big|^2 - \big| \langle o^\sigma | L_x | u^{\sigma^{\prime}} \rangle \big|^2}{E_u^{\sigma^{\prime}} - E_o^\sigma} \notag,		
	\end{align}
	where $\xi$ is the SOC strength, $L_x$ and $L_z$ are
	angular momentum operators, and ${E_u^{\sigma^{\prime}}}$ and $E_o^\sigma$ are the energy levels of unoccupied states $|u^{\sigma^{\prime}} \rangle$ with spin ${\sigma^{\prime}}$ and occupied states $\langle o^\sigma |$ with spin $\sigma$, respectively.
	The term ${\big| \langle o^\sigma | L_z | u^{\sigma^{\prime}} \rangle \big|^2 - \big| \langle o^\sigma | L_x | u^{\sigma^{\prime}} \rangle \big|^2}$ represents the difference of SOC matrix elements for $p$-orbitals, as shown in Supplementary Table III. Therefore, the energy gap ${E_u^{\sigma^{\prime}} - E_o^\sigma}$ plays a key role in contributing to the MAE. Figures.~\ref{2}h, i show the PDOS of $p$-orbitals of Te and I atoms in the MoTeI\textbf{/}In$_2$Se$_3$ heterostructure. For both Te and I atoms, the unoccupied orbitals are primarily located in the spin-up channel, while the occupied orbitals have contributions from both spin channels. Thus, we focus on the energy gap between the unoccupied spin-up $p_x/p_y$ orbitals and the occupied spin-down $p_y/p_x$ orbitals. Additionally, the PDOS shows that under the P$\downarrow$ polarization, both the occupied and unoccupied $p_x/p_y$ orbitals of Te and I atoms shift upward. This significant upward shift in the unoccupied $p_x/p_y$ orbitals increase the overall energy gap ${E_u^{\sigma^{\prime}} - E_o^\sigma}$. Since the SOC matrix element difference between the spin-up (spin-down) occupied $p_x/p_y$ and spin-down (spin-up) unoccupied $p_y/p_x$ states is $-1$, the contribution of $p_x/p_y$ orbital hybridization to MAE increases when FE polarization switches from P↑ to P$\downarrow$. This behavior is consistent with the trends shown in Figs.~\ref{2}d-g.
	
	\begin{figure*}[t]
		\includegraphics[width=1\linewidth]{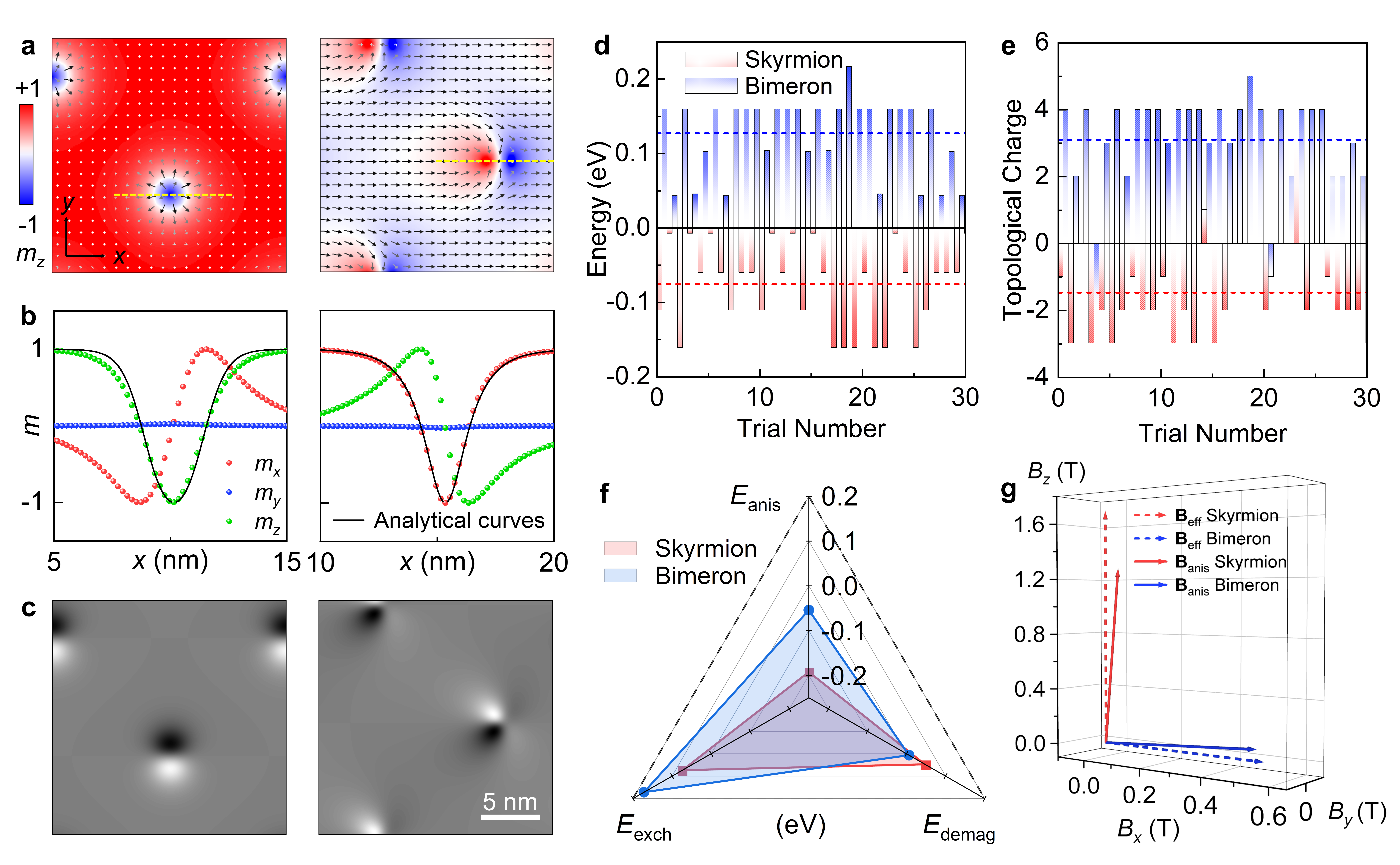}
		\caption{\textbf{Equilibrium States of Skyrmions and Bimerons.} \textbf{a} Spin configurations of equilibrium skyrmions (left column) and bimerons (right column). The yellow dashed lines indicate the direction along the diameter of a skyrmion/bimeron. \textbf{b} The magnetization components $m_x$, $m_y$, and $m_z$ along the diameter direction of equilibrium skyrmions and bimerons. The colored dots represent the simulation results, and the black lines are analytic curves based on Eq.(\ref{3}). \textbf{c} Simulated L-TEM images of equilibrium skyrmions and bimerons. Statistics of thirty sets of \textbf{d} energy and \textbf{e} topological charge data for skyrmion/bimeron relaxed from random magnetization. The dashed lines represent the average values. \textbf{f} Comparison of the average exchange energy $E_{\rm exch}$, anisotropy energy $E_{\rm anis}$, and demagnetization energy $E_{\rm demag}$ for equilibrium skyrmions and bimerons. \textbf{g} The average effective field vectors $\textbf{B}_{\rm eff}$ and anisotropy field vectors $\textbf{B}_{\rm anis}$ for equilibrium skyrmions and bimerons.
		}\label{3} 
	\end{figure*}
	
	\bigskip
	{\textbf{Anisotropy-determined equilibrium states of skyrmions and bimerons.}}
	In this subsection, we investigated the impact of magnetic anisotropy on equilibrium topological spin textures. The initial state of the system is set to a random magnetization (see Supplementary Fig. 5a), followed by relaxation under out-of-plane anisotropy corresponding to P$\downarrow$ and in-plane anisotropy corresponding to P$\uparrow$. The results of the spin configurations are presented in Fig.~\ref{3}a. It can be observed that skyrmions and bimerons can spontaneously form and reach equilibrium states under zero external magnetic field. Out-of-plane anisotropy favors the formation of skyrmions, whereas in-plane anisotropy favors the formation of bimerons. The presence of half-solitons at the boundaries arises from the periodic boundary conditions (PBC). To further confirm the profiles of the skyrmions and bimerons, we extracted the magnetization components $m_x$, $m_y$, and $m_z$ along the yellow dashed line, as illustrated in Fig.~\ref{3}b. The colored dots represent the simulation results, and the black lines are analytical curves fitting with\cite{wang2018theory}
	\begin{align}
		m_{z,x}(x)=2\tan^{-1}\left[\frac{\sinh (d/2w)}{\sinh (x/w)}\right] 
		,\label{3} 
	\end{align}
	where $w$ is the width of the $360^\circ$ domain wall, treated as a fitting parameter, and $d$ is the diameter of the skyrmion/bimeron, which is 2.8 nm for the skyrmion and 2.1 nm for the bimeron. Given that our simulation results agree well with the theoretical formula, we confirmed that the obtained spin configurations are standard skyrmion/bimeron solitons. To better bridge the gap between theoretical results and real experiments, we also presented the simulated Lorentz transmission electron microscopy (L-TEM) images\cite{mccray2021understanding} of the equilibrium skyrmion/bimeron in Fig.~\ref{3}a. In the small defocus limit, assuming full electron-wave processing of the electron beam, the L-TEM contrast originates from the underlying spin configurations. It is characterized by the curl of the magnetization along the beam propagation axis $\hat{\textbf{z}}$, as given by\cite{pollard2017observation}
	\begin{align}
		I({\bf r},\Delta)=1-(\Delta e \mu_0 \lambda t / \hbar)(\nabla\times{\bf m(\bf r)})\cdot\hat{\textbf{z}}
		,\label{4} 
	\end{align}
	where $I$, $\Delta$, $e$, $\mu_0$, $\lambda$, $t$, $\hbar$ are the normalized intensity, degree of defocus, electron charge, vacuum permeability, electron wavelength, magnetic monolayer thickness, and Planck's constant, respectively. Here, a sample tilt of $20^\circ$ was taken to replicate real experiments. These simulated L-TEM images could be compared with previously observed skyrmions and bimerons\cite{yu2024spontaneous} and serve to support the experimental validation of our calculations. It should be mentioned that all equilibrium states obtained in the main text were simulated at 0 K, while we also considered the system at finite temperature (see Supplementary Fig.~5d,e). The results indicate that thermal fluctuations induce local spin perturbations in skyrmions/bimerons, but their topological properties remain unaffected.
	
	\begin{figure*}[t]
		\includegraphics[width=1\linewidth]{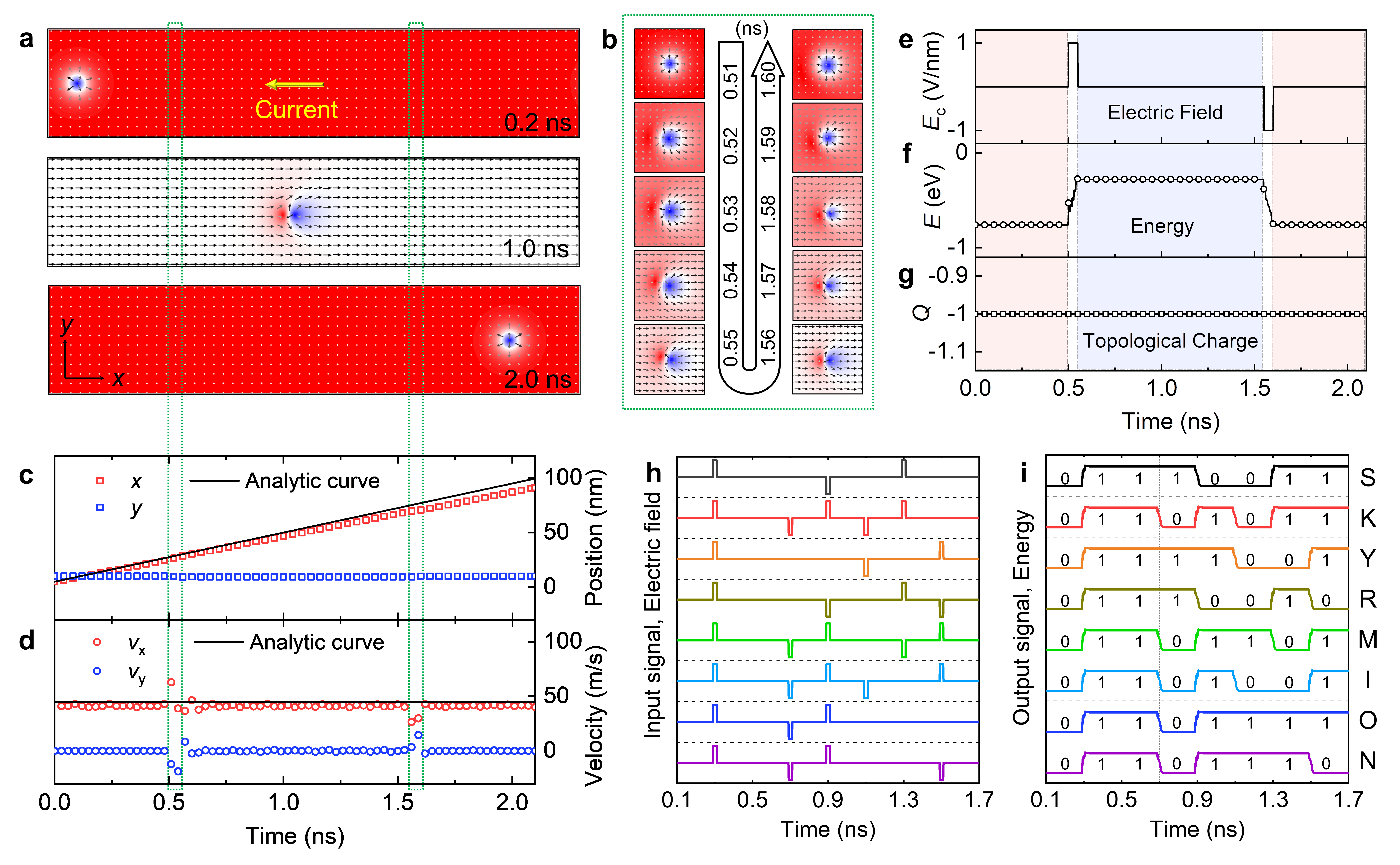}
		\caption{\textbf{Dynamic skyrmion-bimeron transition and its application as an encoder.} \textbf{a} Dynamic snapshots of the moving skyrmion/bimeron in a nanotrack, taken at 0.2 ns, 1.0 ns, and 2.0 ns. \textbf{b} Dynamic snapshots of local spin configuration at the moment of skyrmion-bimeron transitions, taken at 0.51–0.55 ns and 1.56–1.60 ns. Time-dependent variations in \textbf{c} position $(x,y)$ and \textbf{d} velocity $(v_x,v_y)$ during the skyrmion/bimeron motion. The colored dots represent the simulation results, and the black lines are analytic curves based on Eq.(\ref{5}). The green dashed lines mark the moment when the skyrmion-bimeron transitions occur. Time-dependent variations in \textbf{e} electric field $E_{\rm c}$, \textbf{f} total energy $E$, and \textbf{g} topological charge $Q$ during the skyrmion/bimeron motion. A scheme for a dynamic skyrmion-bimeron encoder, with \textbf{h} the electric field as the input signal and  \textbf{i} the total energy as the output signal. The output signal is controlled to represent the binary ASCII code of "SKYRMION."
		}\label{4} 
	\end{figure*}

	As the relaxation process begins from an initially randomized magnetization, the resulting equilibrium states exhibit inherent randomness. That means the skyrmion/bimeron configurations shown in Figs.~\ref{3}a-c represent just one of the many possible equilibrium outcomes. To derive a more generalized conclusion, 30 trial numbers were assigned to 30 sets of independent simulations, each set consisting of two opposite polarization states relaxed from random initial condition, which may still converge to slightly different metastable states due to numerical sensitivity and near-degeneracy. Some typical relaxed configurations of the trials are shown in Supplementary Fig.~5b,c. The statistical analysis is shown in Figs.~\ref{3}d, e, which present the total energy and topological charge, highlighting distinct differences between skyrmions and bimerons despite their inherent randomness. Specifically, the average energy and topological charge are -0.08 eV and -1.5 for skyrmions, and 0.13 eV and 3.1 for bimerons, respectively. These findings suggest that, despite the random variations in individual spin arrangements, the distinction in the topological properties of skyrmions and bimerons remains highly pronounced. Such statistical variation is also consistent with experimental observations, where exact reproduction of identical spin textures is extremely rare, yet their topological and energetic characteristics remain robust. We also noted that the topological charge is reversed in a few instances, this is attributed to the polarity inversion of the skyrmion/bimeron. Figure~\ref{3}f compares the average values of each of the three energy terms for skyrmions and bimerons: exchange energy $E_{\rm exch}$, anisotropy energy $E_{\rm anis}$, and demagnetization energy $E_{\rm demag}$, with the triangular area approximating the total energy. The lower total energy of skyrmions primarily results from the contribution of $E_{\rm anis}$ and $E_{\rm exch}$ terms, whereas bimerons display lower $E_{\rm demag}$. In addition to the energy terms, we also calculated the three-dimensional field vectors, as shown in Fig.~\ref{3}g. It is evident that the effective field $\textbf{B}_{\rm eff}$ points out-of-plane/in-plane for skyrmions/bimerons. Moreover, the direction of the average anisotropy field $\textbf{B}_{\rm anis}$ aligns closely with $\textbf{B}_{\rm eff}$, differing by only a small angle. This also indicates that the equilibrium configurations of skyrmions and bimerons are indeed determined by magnetic anisotropy.

	\bigskip
	{\textbf{Dynamic skyrmion-bimeron transformation in a nanotrack.}}
	Having demonstrated that the out-of-plane/in-plane magnetic anisotropy induced by P$\downarrow$/P$\uparrow$ favors the formation of equilibrium skyrmion/bimeron states, we now examined their dynamic transformation during motion in a nanotrack. A skyrmion is initially positioned at the left end of the nanotrack and begins to move under a continuous in-plane electric current. Electric field pulses are applied at 0.50 ns and 1.55 ns to induce polarization switching in the ferroelectric monolayer, aiming to alter the topological soliton (skyrmion or bimeron) during motion. Figure~\ref{4}a shows dynamic snapshots of the moving spin configuration on the nanotrack at 0.2 ns, 1.0 ns, and 2.0 ns, visually illustrating the back-and-forth switching between skyrmion and bimeron. Snapshots of the local spin configuration during transitions induced by the voltage pulses are shown in Fig.~\ref{4}b. It can be observed that a skyrmion smoothly transitions to a bimeron within as short as 0.05 ns, and vice versa. This fast transformation is attributed to the short duration of the polarization switching assumed in our simulations. In practice, ferroelectric switching typically occurs over nanosecond timescales, depending on the profile and duration of the applied electric field pulse. To assess the robustness of our mechanism under such realistic conditions, we performed additional simulations with a slower polarization modulation. As shown in Supplementary Fig.~6a, even when the transition time is extended from 0.05 ns (used in Figs.~\ref{4}e-g) to 2 ns — a 40-fold increase — the skyrmion–bimeron conversion still proceeds reliably, albeit more gradually. This result confirms the flexibility and dynamical stability of the proposed approach, even under substantially slower, experimentally relevant switching dynamics.
	
	To further analyze the spin dynamics, we extracted the soliton’s position and velocity information, as shown in Figs.~\ref{4}c, d. Although fluctuations in the soliton's velocity occur at the transition moment, the overall process can be regarded as uniform linear motion along the nanotrack. Such robustness in the transition arises from the shared topological charge between skyrmions and bimerons. To support the simulation results, we also perform an analytical analysis based on the modified Thiele equation\cite{thiele1973steady}
	\begin{align}
		\bf{G} \times(\bf{u}-\bf{v})+\pmb{\mathcal{D}}\left(\beta \bf{u}-\alpha \bf{v}\right)=\bf{0}
		,\label{5}
	\end{align}
	where ${\bf G}$ is the gyromagnetic coupling vector, $\pmb{\mathcal{D}}$ is the dissipative force tensor, $\beta$ is the degree of non-adiabaticity, $\alpha$ is the Gilbert damping constant, and $\bf{v}$ and $\bf{u}$ are the velocity of the soliton and the conduction electrons, respectively. In the main text, we set $\alpha=\beta$ to avoid the skyrmion Hall effect, allowing for a more focused study on the skyrmion-bimeron transition. As shown by the solid black line in Fig.~\ref{4}d, the analytic solutions are $v_x\approx45\,\rm{m/s}$ and $v_y=0$ (see Methods for details), which agree well with the simulations. Moreover, we also considered the case of $\alpha\neq\beta$ (see Supplementary Fig.~6b), which exhibits a pronounced skyrmion Hall effect without disrupting the skyrmion-bimeron transition, further demonstrating the robustness of the transition. In addition to position and velocity, we also recorded the variations in electric field $E_{\rm c}$, total energy $E$, and topological charge $Q$ during motion, as shown in Figs.~\ref{4}e-g. The coercive field $E_{\rm c}$ required to switch the FE polarization of In$_2$Se$_3$ is approximately 1 V/nm\cite{bai2024intrinsic}, which is chosen as the peak amplitude for the applied electric field pulse. Notably, as continuous application of the electric field is unnecessary once polarization saturation is achieved, the pulse duration presented here represents the minimum time but can be extended based on practical requirements. By comparing Figs.~\ref{4}e, f, one may observe that the dynamic skyrmion-bimeron transition alters the system energy without changing the topological charge, providing an advantage for the design of skyrmion-bimeron spintronic devices. 
	
	Here, we proposed a scheme for a binary encoder based on the dynamic transformation between skyrmions and bimerons, which is analogous to the racetrack memory. We designated the skyrmion state and bimeron state as binary digits "0" and "1", respectively, enabling repeated transitions between these states to facilitate switching between "0" and "1". The input signal is an electric field that is crucial for determining the spin configuration, while the output signal is a physical quantity (such as energy) that reflects the distinction between the two solitons. The time length of each bit is set to 0.2 ns, and different bits are read out using a clocking system. The variations of the designed sets of eight input and output signals are illustrated in Figs.~\ref{4}h, i. By applying electric field pulses at the appropriate moments, we effectively control the output signal to be eight letters "SKYRMION" in ASCII code, thereby validating the feasibility of this scheme.

	\section*{Discussion}
	In conclusion, we have investigated a reversible transition between skyrmions and bimerons in a MoTeI/In$_2$Se$_3$ multiferroic heterostructure. First-principles calculations revealed that the reversal of ferroelectric polarization in In$_2$Se$_3$ induces a significant shift in the magnetic anisotropy of the MoTeI monolayer from out-of-plane to in-plane orientations, establishing favorable conditions for the formation of skyrmions or bimerons. Micromagnetic simulations further clarified the equilibrium spin textures and dynamic transformations, with statistical analysis and solutions of the Thiele equation confirming both flexibility and stability of the skyrmion-bimeron transition. Building on these findings, we proposed an all-electric-controlled device resembling a racetrack memory and validated its feasibility. Our results presented a promising pathway toward low-power memory applications in future spintronic devices.
	
	Moreover, while our study focuses on a 2D van der Waals heterostructure, the proposed mechanism of electric-field-induced magnetic anisotropy modulation and topological state conversion may be extended to other quantum magnetic materials beyond two dimensions. This includes systems such as hematite\cite{harrison2022route}, ruthenates\cite{wang2018ferroelectrically}, and Heusler compounds\cite{jena2020elliptical}, where tunable magnetic order and spin–orbit interactions offer promising opportunities for electric control of topological spin textures.

	\section*{Methods}
	\textbf{First-principles calculations.}
	All first-principles calculations were performed using the Vienna Ab Initio Simulation Package (VASP) within the framework of DFT\cite{kresse1993ab,kresse1996efficient}. The Perdew-Burke-Ernzerhof (PBE) functional, formulated in the form of generalized gradient approximation (GGA)\cite{perdew1996generalized}, was employed to describe the exchange-correlation energy. Electron-ion interactions were treated using the projector-augmented wave (PAW) pseudopotential method\cite{kresse1999ultrasoft,blochl1994projector}, with a plane wave cutoff energy of 500 eV. To eliminate interactions between adjacent periodic images, a vacuum region of 25 Å was added along the $z$-axis. The $k$-space integration for all calculations was performed using a $15\times15\times1$ $k$-point grid in the 2D Brillouin zone. The criterion of atomic position relaxation was set to $1\times10^{-5}\, \rm{eV/A}$, and the convergence criterion for electron self-consistency was set to $1\times10^{-7}\, \rm{eV}$. To account for the strong electron-electron interactions in localized Mo $d$ orbitals, we employed the ${\rm DFT}+U$ method using Dudarev's approach\cite{dudarev1998electron}, with an effective Hubbard Coulomb interaction parameter, $U_{\rm eff}$, set to 3 eV\cite{wang2023modulating,kang2022reshaped,mishra2021phase,guguchia2018magnetism}. The DFT-D3 method of Grimme was adopted to describe the interlayer vdW interactions\cite{grimme2010consistent,grimme2011effect}. The phonon dispersions were calculated using the finite displacement method as implemented in the Phonopy code\cite{togo2015first}, employing a $4\times4\times1$ supercell. The energy barrier and transition pathway for ferroelectric switching were investigated via the climbing-image nudged elastic band (CI-NEB) approach\cite{henkelman2000climbing,sheppard2012generalized}.
	
	\textbf{Micromagnetic simulations.}
	The micromagnetic simulations were preformed using the GPU-accelerated finite-difference MuMax3 code\cite{vansteenkiste2014design}. Only the magnetic monolayer MoTeI is explicitly modeled and discretized into cells sized $0.1\times0.1 \,\rm{nm}^2$ to maximize computational accuracy. For the static case, the simulation universe is $20\times20 \,\rm{nm}^2$, with PBC applied in both $x$ and $y$ directions, and the energy minimum states are obtained using the conjugate gradient method. For the dynamic case, the simulation universe is $100\times20 \,\rm{nm}^2$, with PBC applied only in the $y$ direction, and calculations are performed using the Dormand–Prince solver.
	
	The average energy density of the system is given by 
	\begin{align}
		&\epsilon=A(\nabla {\bf m})^2+D\left[m_{z}(\nabla \cdot {\bf m})-({\bf m} \cdot \nabla) m_z\right]
		\nonumber\\
		&\ \  -K_{\rm u}({\bf m}\cdot{\hat{\textbf{z}}})^2-\frac{1}{2} \mu_0 M_{\rm s} {\bf m} \cdot {\bf H}_{\rm dm}
		,		
	\end{align}
	where $A$, $D$, and $K_{\rm u}$ are the Heisenberg exchange, interfacial DMI, and magnetic anisotropy constants, respectively. $M_{\rm s}$ is the saturation magnetization and ${\bf H}_{\rm{dm}}$ is the demagnetizing field. The dipolar interaction is always considered in the micromagnetic simulations. Additionally, a term $K_{\rm u}({\bf m}\cdot{\hat{\textbf{x}}})^2$ was incorporated to define an easy axis along the $x$-direction when stabilizing the bimerons. This modeling choice is consistent with the system’s easy-plane anisotropy and does not affect the generality of the results\cite{castro2023skyrmion}. To explore the skyrmion-bimeron transition during the current-induced motion, we numerically solved the Landau-Lifshitz-Gilbert (LLG) equation augmented with a spin-transfer (STT) term $\tau_{\rm STT}$ in Zhang-Li form\cite{zhang2004roles}
	\begin{align}
		{\partial_t {\bf m}}=-\gamma_{0} {\bf m} \times {\bf B}_{\rm eff}+\alpha({\bf m} \times \partial_t {\bf m})+\tau_{\rm STT},
		\\
		\tau_{\rm STT}=u\left( {\bf m} \times \partial_x{\bf m}\times {\bf m}\right) -\beta u\left( {\bf m} \times \partial_x{\bf m}\right),
	\end{align} 
	where $\gamma_{0}$ is the gyromagnetic ratio, and $ {\bf B}_{\rm eff}=-(\delta \epsilon / \delta {\bf m}) /(\mu_0 M_{\rm s})$ is the effective field. The first term of $\tau_{\rm STT}$ is the adiabatic torque and the second term is the non-adiabatic torque. The STT coefficient is given by $u=(g\mu_{\rm B}Pj)/(2eM_{\rm s})$\cite{yamane2016spin}, where $g$, $\mu_{\rm B}$, $P$, and $j$ are the Land\`{e} factor, Bohr magneton, spin polarization factor, and applied current density, respectively. The current density was set to $j=10^{12}\,\rm{A/m^2}$ with $P=0.45$. 
	
	The material parameters used in micromagnetic simulations (Supplementary Table IV) were derived from rigorous DFT calculated parameters (Supplementary Table II) using the following conversion formulas\cite{ma2023strong}: $A={(\sqrt{3}J)}/{(2t)}$, $D={(-\sqrt{3}d_{\Arrowvert})}/{(at)}$, $K_{\rm u}={(2K)}/{(\sqrt{3}a^2t)}$, and $M_{\rm s}={(6\mu_{\rm B})}/{(\sqrt{3}a^2t)}$. Here, $t$ is the thickness of the magnetic monolayer, 3.304 Å for P$\downarrow$ and 3.309 Å for P$\uparrow$, respectively. $a=4.10$ Å is the lattice parameter of the heterostructure for both P$\downarrow$ and P$\uparrow$ states. Polarization switching is modeled by changing the micromagnetic parameters corresponding to the P$\downarrow$ and P$\uparrow$ states, instead of introducing an explicit electric field.
	
	\textbf{Analytical calculation of the Thiele equation.}
	In the modified Thiele equation, the first term describes the Magnus force with ${\bf G}=(0, 0, \mathcal{G})$ and $\mathcal{G}=4\pi Q$. The second term represents the dissipative force where the components $\mathcal{D}_{ij}=\mathcal{D}$ for $(i, j)=(x, x)$ or $(y,y)$ and $\mathcal{D}_{ij}=0$ for otherwise. 
	
	Thus, the general solutions of Eq.(\ref{5}) are\cite{iwasaki2013universal} 
	\begin{align}
		\bf{v}_{\parallel} & =\left(\frac{\beta}{\alpha}+\frac{\alpha-\beta}{\alpha^3(\mathcal{D} / \mathcal{G})^2+\alpha}\right) \bf{u} 
		,
		\\
		\bf{v}_{\perp} & =\frac{(\alpha-\beta)(\mathcal{D} / \mathcal{G})}{\alpha^2(\mathcal{D} / \mathcal{G})^2+1}\left(\hat{z} \times \bf{u}\right)
		,
	\end{align}
	where $\bf{v}_{\parallel}$ and $\bf{v}_{\perp}$ are the components parallel and perpendicular to $\bf{v}$, respectively. When the current injects along the $-x$ direction and $\alpha=\beta$, one can obtain $v_x=u=(g\mu_BPj)/(2eM_{\rm s})$ and $v_y=0$.

	\begin{acknowledgments}
		L.B. acknowledges the support received as a JSPS International Research Fellow at Waseda University. Y.Z. acknowledges support by the National Natural Science Foundation of China (Grant No. 12374123), the Shenzhen Fundamental Research Fund (Grant No. JCYJ20210324120213037), the 2023 SZSTI stable support scheme, and the National Natural Science Foundation of China (Grant No. 12204396). Z.T. acknowledges support by the National Key R\&D Program of China (Grant No. 2023YFB3003004), and the National Natural Science Foundation of China (Grant No. 62376091). 	X.Z. and M.M. acknowledge support by the CREST, the Japan Science and Technology Agency (Grant No. JPMJCR20T1). X.Z. also acknowledges support by the Grants-in-Aid for Scientific Research from JSPS KAKENHI (Grant No. JP25K17939 and No. JP20F20363). M.M. also acknowledges support by the Grants-in-Aid for Scientific Research from JSPS KAKENHI (Grants No. JP25H00611, No. JP24H02231, No. JP23H04522, and No. JP20H00337) and the Waseda University Grant for Special Research Projects (Grant No. 2025C-133).
	\end{acknowledgments}
	
	\newpage

	\bibliographystyle{unsrt}
	\bibliography{Manuscript}
	
\end{document}